\documentstyle[fleqn,12pt,epsf]{article}
\newcommand{\nk}{{\bf k}}
\newcommand{\np}{{\bf p}}
\newcommand{\nq}{{\bf q}}
\newcommand{\nr}{{\bf r}}

\newcommand{\ns}{\mbox{\boldmath{$s$}}}

\newcommand{\nF}{{\bf F}}

\newcommand{\nL}{{\bf L}}

\newcommand{\nP}{{\bf P}}

\newcommand{\nsigma}{\mbox{\boldmath$\sigma$}}

\newcommand{\nOmega}{\mbox{\boldmath$\Omega$}}
\begin{document}
\begin{titlepage}
\thispagestyle{empty}
\mbox{} 
\vspace*{2.5\fill} 

{\Large\bf 
\begin{center}

Final-state interaction and recoil polarization in $(e,e'p)$ 
reactions: comparison with the polarized target case

\end{center}
} 

\vspace{1\fill} 

\begin{center}
{\large
J.E. Amaro$^1$, 
J.A. Caballero$^2$, 
T.W. Donnelly$^3$,
F. Kazemi Tabatabaei$^1$
}
\end{center}

{\small 
\begin{center}
$^1$ {\em Departamento de F\'{\i}sica Moderna, 
          Universidad de Granada,
          Granada 18071, Spain}  \\   
$^2$ {\em Departamento de F\'\i sica At\'omica, Molecular y Nuclear,  
          Universidad de Sevilla, \\ Apdo. 1065, 
          Sevilla 41080, Spain }\\
$^3$ {\em Center for Theoretical Physics, 
          Laboratory for Nuclear Science 
          and Department of Physics, 
          Massachusetts Institute of Technology, 
          Cambridge, MA 02139, USA}\\[2mm]
\end{center}
} 

\kern 1.5cm 
\hrule 
\kern 3mm 

{\small\noindent
{\bf Abstract} 
\vspace{3mm} 

A study of the total cross section for polarized proton knockout in
$(e,e'{\vec p})$ reactions is carried out for the closed-shell nucleus
$^{40}$Ca.  The dependence of FSI effects on polarization observables
viewed as functions of the nucleon polarization angles is analyzed 
and interpreted within the basis of a semi-classical model for the 
orbit of the struck nucleon and trajectory of the ejected nucleon.  
A comparison with the case of a $^{39}$K polarized target and 
unpolarized protons is performed. 
} 

\kern 3mm 
\hrule 

\end{titlepage}
\newpage
\setcounter{page}{1}


\section{Introduction}


In recent years the study of recoil polarization observables in $(e,e'{\vec p})$
and $({\vec e},e'{\vec p})$ reactions has attracted attention both from the
experimental \cite{Woo98,Mal00} and 
theoretical \cite{Kel96,Bof96,Ito97,Ryc99,Kel99,Kel99b,Joh99,Udi00,Kaz03,Kaz04} points of
view. In particular, many of the latest studies have been focused on the 
polarization transfer asymmetry in reactions initiated by polarized 
electrons \cite{Mar02a,Mar02b,Mar03,Vig03}. For these significant effort is being expended 
in investigations of light nuclei, with much of the focus being placed on 
the analysis of the polarization transfer in the 
$^{4}$He\mbox{$(\vec{e},e'\vec{p})$}$^{3}$H reaction which has recently been measured
at MAMI \cite{Die01,Str03}. One of the goals of such studies is to
obtain information from the measured polarization transfer ratios about 
medium modifications of the proton form factor.

In the case of medium-weight nuclei the measurement of recoil nucleon polarization
provides valuable information which is complementary to that obtained with unpolarized
ejectiles \cite{Kel96,Bof96,Fru84,Ras89,Bof93}.  Two experiments have been
performed to date, both for quasielastic kinematics.  In one of them
\cite{Mal00}, using polarized electrons, the polarization transfer $P'_l$,
$P'_s$ and ratio were measured for proton knockout from the various shells of
$^{16}$O and high momentum transfer $q=1$ GeV/c.  In the other \cite{Woo98},
with unpolarized electrons, the induced normal polarization $P_n$ was
determined for the two shells of $^{12}$C, and a lower value of the momentum
transfer, $q\simeq 760$ MeV/c.  The agreements reached with the experimental data
using different Distorted-Wave Impulse Approximation (DWIA) models
\cite{Ryc99,Joh99,Udi00,Kaz03,Kaz04,Mar03} make these studies promising. 
The polarization observables provide a clean
tool to study reaction mechanisms, nuclear structure details and
electromagnetic properties of the system, and to disentangle the
various theoretical ingredients contained in each model,
since, being ratios of response functions, no ambiguities arise due 
to spectroscopic factors that have to be fitted to data. 
In particular, note that adequate modeling of the final-state interaction 
(FSI) is essential for a proper description of so-called Time-Reversal Odd 
(TRO) observables~\cite{Ras89} such as the induced polarization, since they 
would be exactly zero in absence of FSI.

In this paper we are mainly interested in the polarization observables induced
by {\em unpolarized} electrons, specifically reactions of the type 
$(e,e'{\vec N})$ --- in this work only the case $N=p$ will be considered, but
the formalism is also applicable to the case $N=n$. 
These are also TRO observables and, as stated above, they only arise in the
presence of FSI, in contrast to the polarization transfer observables produced
by polarized electrons, which are Time-Reversal Even (TRE) and are 
already different from zero in the Plane-Wave
Impulse Approximation (PWIA).  When the initial electron and target are
unpolarized, the electron can hit with equal probability nucleons having all
spin orientations at different places along its orbit.  In absence of FSI
these nucleons leave the nucleus as plane waves with the same amplitude, hence
carrying no net induced polarization.  

The mechanisms by which the FSI induce
recoil polarization are mainly i) the absorption due to the imaginary part of
the optical potential and ii) the spin-orbit interaction.  

The first mechanism is similar to the one found in quasielastic proton
scattering (News polarization \cite{New53,New58}, Maris effect \cite{Jac76}).
In these reactions the dominance of one side of the nucleus (with respect to
the final momentum of the nucleon, $\np'$) emphasizes a definite orientation
of the initial orbit, and then also favors a definite orientation of the
nucleon spin.  An analogous effect was independently predicted for $(e,e'p)$
reactions from polarized nuclei in \cite{Ama99}. Therein the effects of
FSI on the cross section were shown to depend strongly on the nuclear
orientation.  This result was interpreted by a semi-classical picture of the
reaction: for a given polarized shell and given missing momentum,
$\np=\np'-\nq$, the most probable value of the nucleon position can be
determined, and from that one can evaluate the amount of nuclear matter that
is crossed by the nucleon; this in turn can be related with the strength of
absorption due to the imaginary part of the optical potential.

The results found in \cite{Ama99} and later extended to polarized
electrons in \cite{Ama02} provide a clear geometric picture of
these reactions on polarized nuclei.  One of the
aims of the present paper is to investigate the degree to which this picture can be
extended to the case of induced nucleon polarization in $(e,e'{\vec p})$
reactions with unpolarized targets.  Results for the induced
polarization in such reactions have already been interpreted in terms
of the Maris effect in \cite{Woo98,Kel96}.  In this paper we go in some depth
into this interpretation of nucleon recoil polarization in the light
of the results already found for polarized nuclei. We depend on the fact 
that effects beyond the
impulse approximation arising from two-body meson exchange currents do not
affect the present results for the $(e,e'p)$ cross section for low
missing momentum and quasielastic kinematics \cite{Kaz03,Kaz04},
which is the region where the semi-classical picture can safely be
applied.  In particular, in this paper we want to bring out the
differences found between observables which occur for polarizations 
in the initial state (target) and in the final state (recoil nucleon), 
and to highlight the
advantages of one reaction over the other, as this can help to guide
future polarization experiments. 

In this sense the present study is complementary to work in progress 
focused on the more general case of polarization observables which 
occur when {\em both} target and
ejectile are polarized \cite{Cab04}.  Analyses of the complete
set of polarization observables in $(e,e'p)$ only exist for the case
of light nuclei, in particular deuterium \cite{Are04}--\cite{Are92,Dmi89}, but 
not for medium-weight nuclei.  In the present paper we apply
the DWIA model of  \cite{Kaz03,Kaz04,Ama03b} to describe proton
knockout from the $d_{3/2}$ shell of $^{40}$Ca for various
recoil polarization directions, and compare the results with 
those found for the model
of \cite{Ama99,Ama02,Ama98b} of proton knockout from the same
shell of polarized $^{39}$K.

The structure of the paper is as follows: in Sec.~2
we briefly present the general formalism for $(e,e'{\vec p})$ recoil 
polarization reactions which will serve as a basis for our DWIA model. 
In Sec.~3 we give results for the cross section for different polarizations 
of the proton and interpret these using the semi-classical picture of 
nucleon orbit. Connections are made there to the case
of a polarized target. We also present results for other polarization
observables --- in particular the separated response functions.
Finally the conclusions are presented in Sec.~4.


\section{Formalism for $(e,e'\vec{p})$ reactions}


In this section we present the basic formalism for $(e,e'\vec{p})$ reactions 
on which the DWIA results of the next section  are based.  
We closely follow the
pioneering work on polarization observables developed in
\cite{Ras89,Pic87,Pic89}. For brevity here we just give the basic required 
expressions; more details about our model can be found in our
previous work \cite{Kaz03,Maz02} and references therein.  

The kinematics of
the reaction are shown in Fig.~1.  An electron with four-momentum
$K^\mu_e=(\epsilon_e,\nk_e)$ scatters off a nucleus $|A\rangle$ to final
four-momentum $K'_e{}^\mu=(\epsilon'_e,\nk'_e)$, and is detected in coincidence
with a proton with momentum $\np'$. The residual nucleus is left in the
discrete state $|B\rangle$. The energy transfer is
$\omega=\epsilon_e-\epsilon'_e$.  The proton polarization is measured in the
direction $\vec{s}$. In this work we express the vector components in the Lab
coordinate system where the $z$-axis points along the $\nq$ direction, with
$\nq=\nk_e-\nk'_e$ the momentum transfer. The $x$-axis points along the
transverse component of $\nk_e$ (perpendicular to $\nq$), together
with $z$ defining the scattering plane. The $y$-axis is perpendicular to that plane.
The final momentum $\np'$ of the proton with polar angles $(\theta',\phi')$
together with the momentum transfer determines the reaction plane, which
forms an angle $\phi'$ with the scattering plane.  Experimentally the nucleon
polarization components are measured with respect to the barycentric system,
which we also show in Fig.~1 for future reference.  The two directions
longitudinal ($\vec{l}$) and sideways ($\vec{t}$) lie in the reaction plane,
while the normal direction ($\vec{n}$) is perpendicular to it.

\begin{figure}[tp]
\begin{center}
\leavevmode
\def\epsfsize#1#2{0.9#1}
\epsfbox[190 480 425 700]{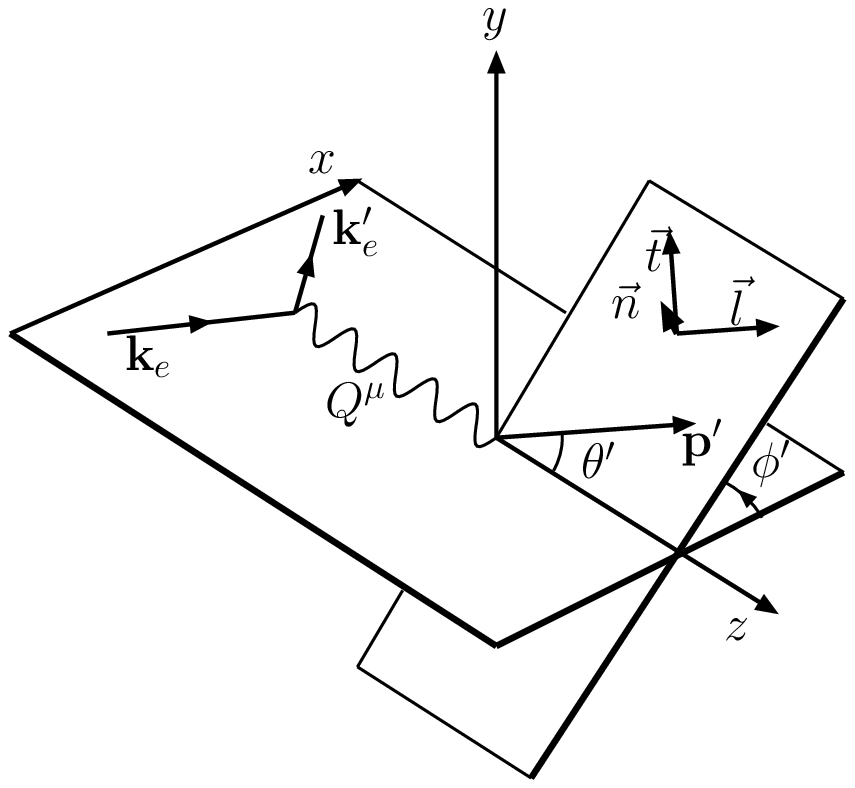}
\caption{Kinematics and coordinate system used in this work.
The $x$-$z$ coordinates span the scattering plane, while the 
ejected nucleon momentum $\np'$ together with the momentum
transfer ($z$ direction) determines the reaction plane.
The polarization vector $\vec{s}$ of the nucleon is often 
referred to the barycentric coordinate system $(\vec{l},\vec{t},\vec{n})$, 
also shown in the figure, although for the results in the present work we
use the Lab system, determined by the $(x,y,z)$ coordinates. 
}

\end{center}
\end{figure}

The $(e,e'{\vec p})$ cross section for protons polarized in the
$\vec{s}$ direction may be written as 
\begin{equation} 
\label{sigma}
\Sigma\equiv\frac{d\sigma}{d\epsilon'_e d\Omega'_e d\Omega'}
=K\sigma_M\left(v_LR^L+v_TR^T+v_{TL}R^{TL}+v_{TT}R^{TT}\right),
\end{equation}
where $\Omega'_e$ and $\Omega'$ are the solid angles of the 
final electron and proton, respectively, $\sigma_M$ is the Mott cross section and 
$K=m_N p'/(2\pi\hbar)^3$, with $m_N$ the nucleon mass. The 
electron kinematical factors $v_K$ are given by
\begin{eqnarray}
v_L = \frac{Q^4}{q^2} &, & v_T = \tan^2\frac\theta2-\frac{Q^2}{2q^2}
\\ v_{TT} = \frac{Q^2}{2q^2} &,& v_{TL} =
\frac{1}{\sqrt{2}}\frac{Q^2}{q^2}
\sqrt{\tan^2\frac\theta2-\frac{Q^2}{q^2}},
\end{eqnarray}
with $Q^2=\omega^2-q^2<0$. Finally the exclusive
nuclear responses are the following components 
\begin{eqnarray} 
R^L = W^{00} 
&,&
R^T=W^{xx}+W^{yy}  
\label{r1}\\
R^{TL}=\sqrt{2}\left(W^{0x}+W^{x0}\right) 
&,&
R^{TT}=W^{yy}-W^{xx}
\label{r2}
\end{eqnarray}
of the hadronic tensor
\begin{equation}
W^{\mu\nu}=\frac{1}{K}\sum_{M_B}\overline{\sum_{M_A}}
\langle \np'\vec{s},B|J^\mu(\nq)|A\rangle^*
\langle \np'\vec{s},B|J^\nu(\nq)|A\rangle.
\end{equation}
Here we average over third components of the angular momentum of the 
target ground state $|A\rangle$
and sum over undetected orientations of the daughter nucleus in the
(asymptotic) discrete state $|B\rangle$, for which we neglect recoil.  The
asymptotic nucleon state $|\np'\vec{s}\rangle$, with spin in the $\vec{s}$
direction, is computed in DWIA as a solution of the Schr\"odinger equation
with an optical potential.  In the particular case of the results for
$^{40}$Ca of the next section, we use the potential parameterization of
Schwandt \cite{Sch82}, the same considered in \cite{Ama99} since we want
to compare with the results for polarized nuclei using the same interaction.
The present model accounts for relativistic corrections that have proven to be
successful in describing intermediate-energy inclusive and exclusive
electron scattering observables in the region of the quasielastic peak
\cite{Udi99,Ama96a,Ama98a,Ama02c,Ama03}.  First we use relativistic kinematics
for the nucleons, as these are essential to have a correct description of the
position and width of the quasielastic peak.  Second we use the so-called
semi-relativistic form of the nuclear (one-body) electromagnetic current
operator, $J^\mu(\nq)$, obtained as an expansion in powers of $\np/m_N$, the
missing momentum over the nucleon mass, maintaining the exact dependence on the
energy-momentum transfer.  The resulting semi-relativistic DWIA model (SR-DWIA) 
was applied to
$(e,e'p)$ reactions from polarized nuclei in \cite{Ama99,Ama02,Ama98b}, 
to recoil polarization observables in \cite{Kaz03,Kaz04}, and to the
unpolarized reaction in \cite{Ama03b,Maz02,Udi99}. Effects beyond the impulse
approximation due to meson-exchange currents (MEC) were discussed in
\cite{Kaz03,Kaz04,Ama03b} --- it can be shown that they do not affect
the results of the present work significantly.

The response functions in Eqs.~(\ref{r1},\ref{r2}) depend on the emission
$(\theta',\phi')$ and
polarization $(\theta_s,\phi_s)$ angles, 
apart from the kinematics. It is usual to extract the 
dependence on the  azimuthal emission angle  $\phi'=\phi$, where 
$(\theta,\phi)$ are the polar angles of the missing momentum $\np=\np'-\nq$.
In this way one writes
\begin{eqnarray}
R^L & = & W^L \label{RL} \\
R^T & = & W^T \\
R^{TL} & = &  W^{TL}\cos\phi+\widetilde{W}^{TL}\sin\phi \\
R^{TT} & = &  W^{TL}\cos2\phi+\widetilde{W}^{TL}\sin2\phi, \label{RTT} 
\end{eqnarray}
where the dependence of the six response functions $W^K$ (with $K=L,T,TL,TT$)
and $\widetilde{W}^K$ (with $K=TL,TT$) on the azimuthal angles is now only
via their difference $\Delta\phi=\phi-\phi_s$.  In order also to display
explicitly the spin-dependence $(\vec{s})$ these response functions are
expanded as sums of spin-scalar $(0)$ plus spin-vector ($V$) response
functions
\begin{eqnarray}
W^K &=& W^K_0+W^K_V = \frac12 W^K_{\rm unpol}+W^K_n s_n   
\kern 1cm  K=L,T,TL,TT 
\label{WK}\\
\widetilde{W}^K &=& \widetilde{W}^K_V = W^K_l s_l + W^K_t s_t
\kern 2cm  K=TL,TT
\label{WKtilde}
\end{eqnarray}
where the scalar responses $W^K_0$ are precisely one-half of the corresponding
unpolarized response (since the sum over two possible final spin orientations
of the nucleon cancels the factor 1/2). The vector responses
have the form of a scalar product and have been written as a linear
combination of the spin components in the barycentric system, $(s_l,s_t,s_n)$,
where the coefficients in this expansion, $W^K_i$,
 are the so-called polarized reduced
response functions.
We can see from Eq.~(\ref{WK}) that in the case of the responses without a tilde,
$W^K_V=W^K_n s_n$, and only the normal component of the spin enters.  The tilde
responses, Eq.~(\ref{WKtilde}), are purely spin-vector; hence they are zero in
absence of polarization and they have no normal component, {\it i.e.}, only the
components of the spin in the reaction plane contribute.  Expressions for the
reduced response functions, $W^K_i$, using the present DWIA model were
presented in \cite{Kaz03} as a general multipole expansion,
and in \cite{Maz02} for the case of the unpolarized ones, $W^K_{\rm unpol}$.

Using the above-defined sets of response functions, the cross section
in Eq.~(\ref{sigma}) can be expanded as the sum of spin-scalar plus spin-vector 
parts
\begin{equation}
\Sigma= \Sigma_0+\Sigma_V=\frac12\Sigma_{\rm unpol}+\Sigma_V, 
\end{equation}
where $\Sigma_{\rm unpol}=2\Sigma_0$ is the unpolarized cross section,
\begin{equation}
\Sigma_{0}=
K\sigma_M
\left(
  v_LW^L_0 + v_TW^T_0 + v_{TL}W^{TL}_0\cos\phi + v_{TT}W^{TT}_0 \cos2\phi
\right)
\end{equation}
and the spin-vector part can be written as a scalar product, {\it i.e},  
 in the barycentric system
\begin{equation}
\Sigma_V= \Sigma_l s_l + \Sigma_t s_t + \Sigma_n s_n
\end{equation}
with 
\begin{eqnarray}
\Sigma_n
&=& 
K\sigma_M 
\left(
  v_LW^L_n+v_TW^T_n+v_{TL}\cos\phi W^{TL}_n +v_{TT}\cos2\phi W^{TT}_n
\right)
\label{sigma_n}
\\
\Sigma_l
&=&
K\sigma_M
\left( v_{TL}\sin\phi W^{TL}_l +v_{TT}\sin2\phi W^{TT}_l
\right)
\label{sigma_l}
\\
\Sigma_t
&=&
K\sigma_M
\left( v_{TL}\sin\phi W^{TL}_t +v_{TT}\sin2\phi W^{TT}_t
\right).
\label{sigma_t}
\end{eqnarray}
Finally in terms of the polarization asymmetries we can write
\begin{equation} \label{sigma-P}
\Sigma =\frac12
\Sigma_{unpol}\left(1+\nP\cdot\vec{s}\right),
\end{equation}
where the induced polarization asymmetry vector, $\nP$, 
has been introduced  as the quotient between
the  spin-vector components and the spin-scalar cross section. 
\begin{equation}
P_i = \Sigma_i/\Sigma_0, \kern 1cm i=n,l,t
\end{equation}
The component of the polarization in
a given direction $\vec{s}$ can be obtained in a coincidence experiment as an
asymmetry by measuring the number $N(\vec{s})$ of protons polarized in the
direction $\vec{s}$ minus the number $N(-\vec{s})$ of protons polarized in the $-\vec{s}$
direction, divided by the sum
\begin{equation}
\nP\cdot\vec{s} = \frac{N(\vec{s})-N(-\vec{s})}{N(\vec{s})+N(-\vec{s})}
\end{equation}
and in this way in experimental studies one can minimize systematic 
errors contained in the separate polarized cross sections. 

The roles of the various ingredients of our model (FSI, MEC, {\it etc.}) in the
induced polarization components were analyzed in \cite{Kaz03,Ama03b} for
selected quasielastic kinematics.  The MEC effects were found to be negligible for
missing momentum less than the Fermi momentum ($\sim 240$ MeV/c), thereby
corroborating our contention that the impulse approximation is appropriate for analyzing
these observables under this kinematics.  

In the following we show results for the cross section and related
observables as functions of the missing momentum for
various polarization directions.


\section{Results}


\subsection{Cross section and semi-classical picture}

In \cite{Ama99} the total cross section for $(e,e'p)$ reactions
from {\em polarized} nuclei was computed using the present SR-DWIA
model and the results were analyzed in the light of a
semi-classical model of the struck nucleon's orbit and mean free path of the
final-state ejected nucleon. The analysis was later 
successfully extended to the case of
polarized electrons \cite{Ama02}. In \cite{Ama99} different FSI
effects were found for different initial polarization directions of
the nucleus. This polarization dependence was analyzed within the context
of the afore-mentioned semi-classical picture of the nucleon orbit. One of the
goals of this section is to show the connections of the polarized
nucleus case ({\it i.e.}, polarization of the initial state) with the case of
nucleon recoil polarization (final state) and with what sometimes has
been called the ``Maris effect'' in the literature
\cite{Woo98,Kel96,Udi00}.  We are interested in emphasizing the
differences between the two reactions with respect to the physical
information that can be obtained, and
the advantages of one over the other from the theoretical point of
view.

\begin{figure}[tp]
\begin{center}
\leavevmode
\def\epsfsize#1#2{0.8#1}
\epsfbox[50 177 540 765]{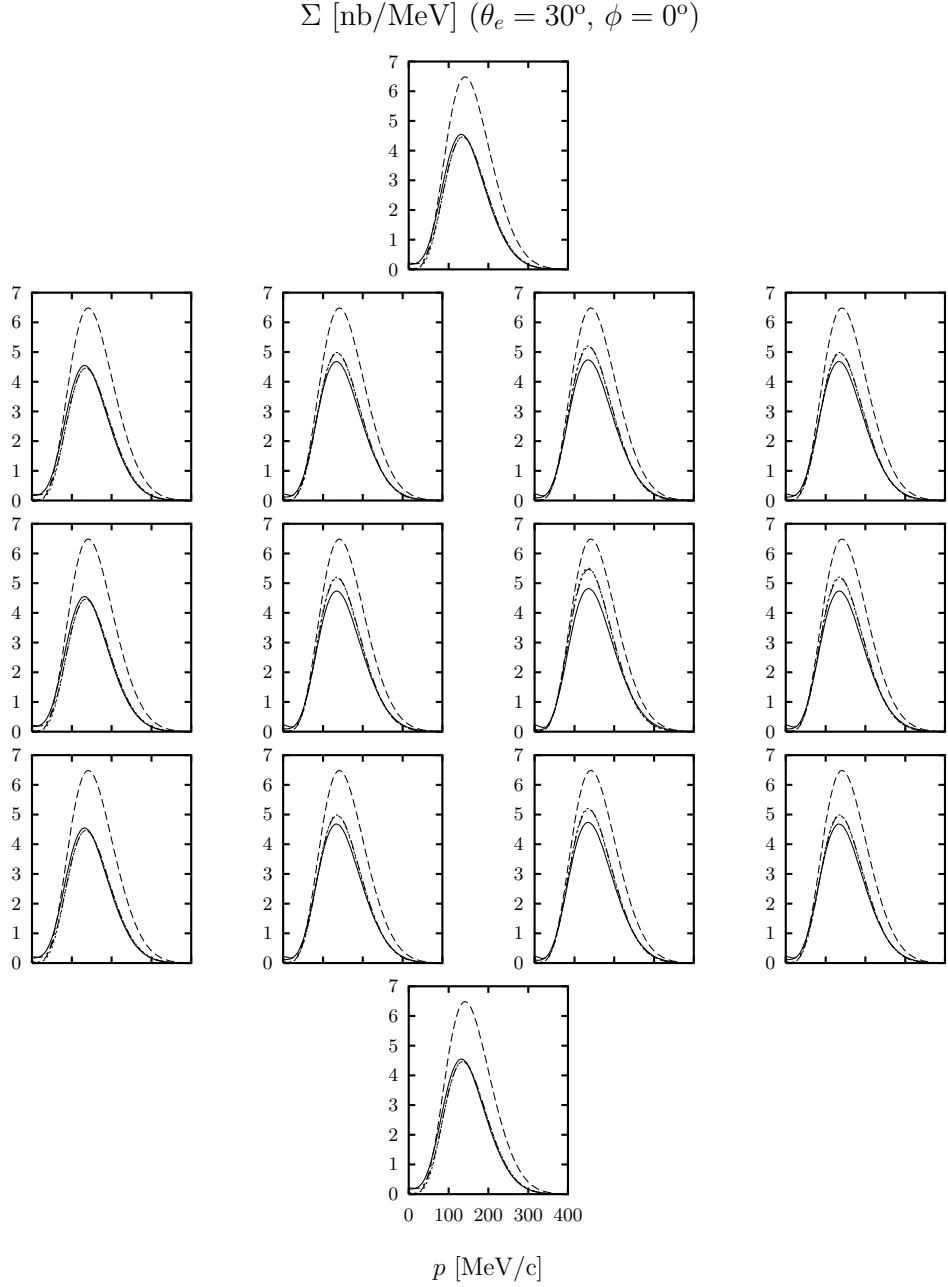}
\caption{Proton knockout cross section from the $d_{3/2}$ shell of
$^{40}$Ca for $q=500$ MeV/c and $\omega=133.5$ MeV.  Each panel
corresponds to a different orientation of the final proton spin
$\vec{s}$. From up to down, $\theta_s=0, 45^{\rm o}$, 90$^{\rm o}$,
135$^{\rm o}$, 180$^{\rm o}$.  From left to right, for
$\Delta\phi\equiv \phi-\phi_s= 0,45^{\rm o}$, 90$^{\rm o}$, and
135$^{\rm o}$. Dashed lines: PWIA. The rest of the curves are from DWIA
calculations with the full optical potential (solid), with $V_{ls}=0$
(dot-dashed), and with Re$V_C=V_{ls}=0$ (dotted).}

\end{center}
\end{figure}

\begin{figure}[tp]
\begin{center}
\leavevmode
\def\epsfsize#1#2{0.8#1}
\epsfbox[50 397 540 765]{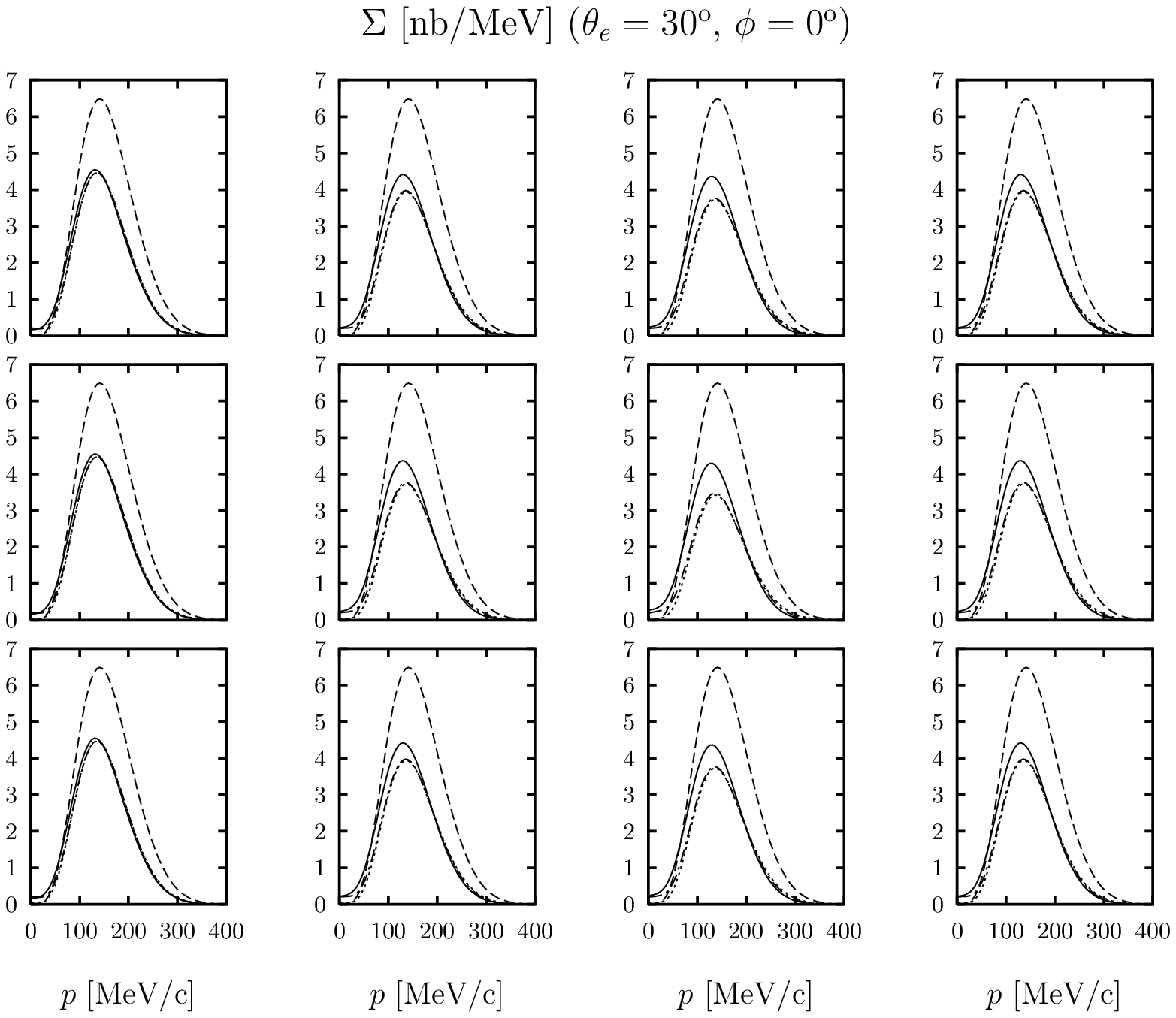}
\caption{As for Fig.~10 for spin angles, from up to down,
$\theta_s=45^{\rm o}$, 90$^{\rm o}$, 135$^{\rm o}$, and 180$^{\rm o}$ and
from left to right, $\Delta\phi=180^{\rm o}$, 225$^{\rm o}$, 270$^{\rm
o}$, and 315$^{\rm o}$.  }
\end{center}
\end{figure}

In order to proceed with this study it is convenient to express the recoil
polarization observables in the Lab system $(x,y,z)$ instead of the
barycentric coordinates $(l,t,n)$ used in the experimental studies. The latter
are relative to the path of the final proton, and hence they depend on the
emission angles, while the former are fixed with respect to the initial
state, and therefore do not depend on the emission angles. Accordingly the Lab system is
appropriate when one wishes to compare with the polarized target case and 
to relate the results
with the corresponding nucleon orbit in the ``initial'' state.

Fixing on an explicit case for discussion, in Figs.~2 and 3 we show the cross
section for proton knockout from the $d_{3/2}$ shell of $^{40}$Ca for
azimuthal angle $\phi=0$.  The various panels correspond to different final
polarization angles, $\theta_s$ and $\Delta\phi\equiv\phi-\phi_s$ for the
proton.  Specifically, in Fig.~2 we show results for angles, from up to down
panels, $\theta_s=0, 45^{\rm o}$, 90$^{\rm o}$, 135$^{\rm o}$, 180$^{\rm o}$,
and, from left to right, for $\Delta\phi=0,45^{\rm o}$, 90$^{\rm o}$,
135$^{\rm o}$.  Thus the results of Fig.~2 are a plot of the cross section for
the ``occidental'' hemisphere in the space of polarization angles, including the
north and south poles.  The second ``oriental'' hemisphere is displayed in
Fig.~3, where the chosen angles are, from up to down, $\theta_s=45^{\rm o}$,
90$^{\rm o}$, 135$^{\rm o}$, and 180$^{\rm}$, and, from left to right,
$\Delta\phi=180^{\rm o}$, 225$^{\rm o}$, 270$^{\rm o}$, 315$^{\rm o}$.  Note
that in the figures the proton exits with $\phi=0$. Hence
$\Delta\phi=-\phi_s$ for the present case.

The $^{40}$Ca nucleus has been chosen in order to make direct comparisons
with the $^{39}$K polarized target considered in \cite{Ama99},
where the initial state was described as a hole in the $d_{3/2}$ shell
of the $^{40}$Ca core. The kinematics $q=500$ MeV/c, $\omega$=133.5
MeV, corresponding to the quasielastic peak, and $\theta_e=30^{\rm o}$, 
have also been chosen as in \cite{Ama99}, where an expanded plot
similar to Figs.~2 and 3 showing results for all of the nuclear
polarization directions was also presented. This allows us to make
direct comparisons with the present case, taking into account that
the two sets of results correspond to different processes and the absolute values of
cross sections are not the same, since there exist
geometrical factors which depend on the polarization conditions. In
particular, in \cite{Ama99} the case of knockout from the $d_{3/2}$
shell of $^{39}$K, and two holes in the final state coupled to $J_B=0$
was considered.  In \cite{Ama98b}, Eq. (82), it was demonstrated that the
corresponding nuclear cross section is related to the one of a single
polarized particle in the $d_{3/2}$ shell, by a factor 1/2 for this
particular case\footnote{In \cite{Ama99} different units (fm$^3$) were
employed for the cross section and the relation 1 nb/MeV$\simeq
20\times 10^{-6}$fm$^3$ must be applied before comparisons can easily be made. }.  

To begin, we have checked the consistence of the calculation
of \cite{Ama99} with the present results. Note that for the $\phi=0$
case considered in Figs.~2 and 3
 the induced polarization has only a normal ($n$) component; see
Eqs.~(\ref{sigma_n}--\ref{sigma_t}). 
 Hence for $\phi_s=\Delta\phi=0$ (north and
south poles and left-hand panels in Fig.~2) or $\Delta\phi=180^{\rm o}$
(left-hand panels in Fig.~3) the induced polarization is zero and the
$^{40}$Ca$(e,e'p)$ unpolarized cross section must be recovered after
multiplying by a factor of 2; see Eq.~(\ref{sigma-P}).  Alternatively, note also
that in PWIA (dashed lines) there is no induced polarization, and thus the
dashed curves in all panels are identical. 

Let us start the comparisons by restricting our attention 
to the PWIA results. By inspection in Fig.~2 we read a
cross section at the maximum  of roughly
$\Sigma\simeq 6.5$ nb/MeV. Hence the unpolarized cross section in PWIA is
(see footnote)
\begin{equation}
\Sigma_{unpol}=2\Sigma
\simeq 13.0\, {\rm nb/MeV} 
\simeq 260\times 10^{-6}\,{\rm fm}^3.
\end{equation}
Since there are four particles in the $d_{3/2}$ shell, 
the unpolarized cross section per particle is
\begin{equation}\label{d32}
\frac{\Sigma_{unpol}}{4} \simeq 65 \times 10^{-6}\,{\rm fm}^3.
\end{equation}
This number can be directly compared with the unpolarized cross 
section of $^{39}$K  in PWIA, displayed with dashed lines 
in Fig.~3 of \cite{Ama99},
which at the maximum is roughly
$\Sigma_{unpol}\simeq 32.5 \times 10^{-6}\,{\rm fm}^3.$
As stated earlier, this differs by a factor 1/2 from the unpolarized cross
section for a particle, namely, $\simeq 65/2 \times 10^{-6}\,{\rm fm}^3$, 
in accordance with Eq.~(\ref{d32}).

Next we discuss the effects that the FSI have on the nucleon
polarization.  These can be seen in Figs.~2 and 3 where we display
the cross section from our DWIA model, showing the contribution of
different terms of the optical potential
\begin{equation}
V_{\rm opt}(r)= V_C(r)+V_{ls}(r)\nsigma\cdot\nL
= U_C(r)+iW_C(r)+\left[U_{ls}(r)+iW_{ls}(r)\right]\nsigma\cdot\nL.
\end{equation}
With solid lines we show the  total DWIA calculation using the full potential.  
To be compared with this, with
dot-dashed lines we show results after setting to zero the spin-orbit
part of the optical potential ($V_{ls}=0$), while the dotted lines
include only the imaginary part of the central optical potential
({\it i.e.}, $V_{ls}=0$, $U_C=0$). Finally the dashed lines do not include
FSI (PWIA). Remember that the PWIA cross section (dashed lines) is the
same for all of the polarizations, in accordance with the fact that there
is no induced polarization in the absence of FSI.  

First, note that the
full FSI (compare the dashed with the solid lines) produces for all 
polarizations a reduction of the cross section in the region around
the maximum of the momentum distribution for the chosen shell. Second, the
magnitude of this reduction depends smoothly on the polarization
direction and varies slowly throughout the various panels shown in the
figures.  Looking at a given pair of panels corresponding to opposite
nucleon polarizations we see that in general the total cross section
is different for the two polarizations.  Under these conditions a net
nucleon polarization component is generated along the given direction.
The exception arises for the cases $\Delta\phi=0, 180^{\rm o}$
spanning the $xz$ plane (left-hand panels in the figures), since the normal
direction, which is in fact the direction of the induced polarization
vector $\nP$ for $\phi=0$, has a zero component in the $xz$ plane.
Note also that the FSI reduction is in general more pronounced for the
polarizations of Fig.~3 than for Fig.~2; this fact is related to the
Maris effect and will be explained below. This determines the sign of
the polarization component in a given direction.

Concerning the various ingredients of the FSI, we see that the
imaginary part of the central optical potential, Im$V_C$ (dotted
lines), is the main one responsible for the reduction of the cross section
due to the absorption effect.  Note that this reduction is again
stronger for the polarizations of Fig.~3 than for Fig.~2 (except for the
$\Delta\phi=0, 180^{\rm o}$ cases which do not depend on
polarization). 
A larger reduction (around 50\%) is found for the
polarization $\theta_s=90^{\rm o}$, $\Delta\phi=270^{\rm o}$ in 
Fig.~3, while it is smaller ($\sim 20\%$) for the
opposite polarization $\theta_s=90^{\rm o}$, $\Delta\phi=90^{\rm o}$
in Fig.~2. If we also add the real part of the central optical
potential, Re$V_C$ (dot-dashed lines), producing only scattering, the
cross section is almost unchanged.  Finally, the inclusion of the
spin-orbit part of the potential, $V_{ls}$ (solid lines), produces an
additional decrease (increase) of the cross sections of Fig.~2
(Fig.~3). This reduces the difference between cross sections for
opposite polarization directions, {\it i.e.}, the spin-orbit interaction produces an
effect that goes towards reducing the induced polarization.

Next we will focus our analysis on the FSI produced exclusively by the
imaginary part of the optical potential (dotted lines in Figs.~2 and
3). Thus we disregard for the moment the effect of the spin-orbit part
of the interaction which also affects the induced polarization, and
which will be considered later.

\begin{figure}[th]
\begin{center}
\leavevmode
\def\epsfsize#1#2{0.8#1}
\epsfbox{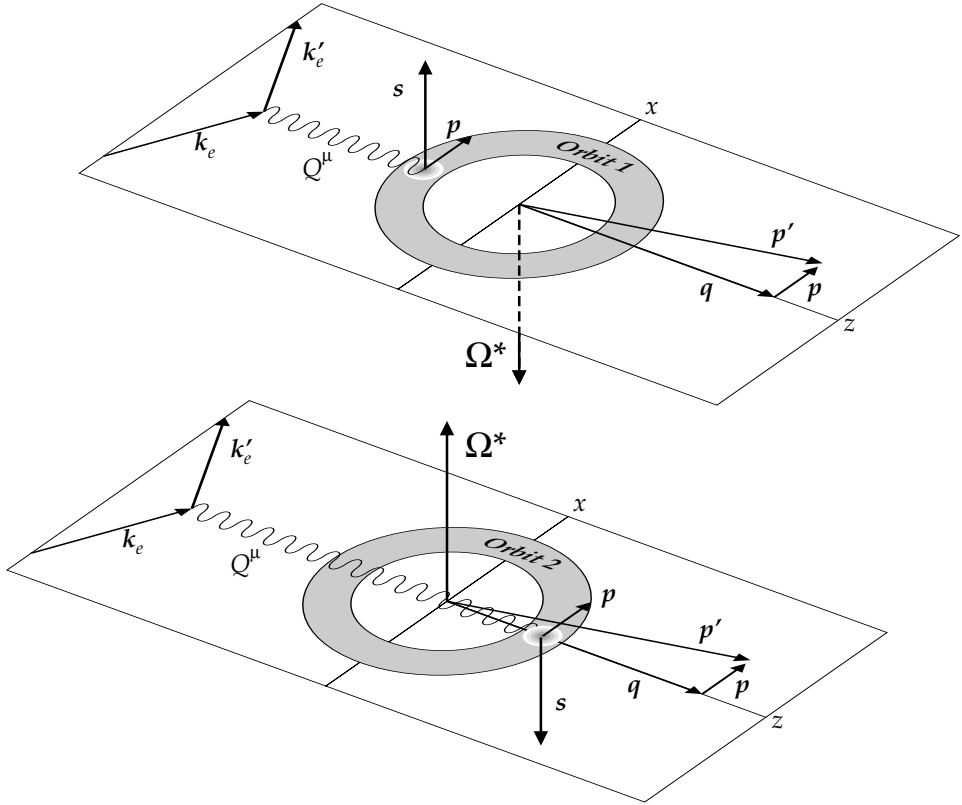}
\caption{Semi-classical picture for opposite polarizations of 
  the nucleon in a $d_{3/2}$ orbit, for identical
  kinematical conditions. The expected position of the proton is indicated
  with a shaded ball.}
\end{center}
\end{figure}

The impact of the FSI as a function of the polarization seen in
Figs.~2 and 3 is qualitatively similar to what was found in the case
of polarized $^{39}$K in \cite{Ama99}, where large absorption effects
for some polarizations and small for the opposite were observed.  These
properties of FSI for polarized nuclei were explained in \cite{Ama99}
within the context of the semi-classical concept of nucleon orbit.  Here
we briefly discuss how that model can describe the present case of
induced nucleon polarization.  Focusing on the $d_{3/2}$ case, we
consider a particle in that orbit, initially polarized in the
direction $\nOmega^*$ (Fig.~4). Here the polarization of the orbit is
the direction of the total angular momentum obtained as the vector sum
of spin plus orbital angular momentum.  In the jack-knifed case of a
$d_{3/2}$ wave one expects that the spin and orbital angular momentum
be opposite.  The spatial distribution of the proton can be pictured
as a torus-like distribution, where the proton is rotating in a
counter-clockwise sense with respect to the polarization direction of the
orbit $\nOmega^*$ (see Fig.~4).  In fact the (local) expectation value
of momentum at a position $\nr$ inside the $d_{3/2}$ orbit is given in
the semi-classical model by
\begin{equation}
\np(\nr) = \frac{1+\sin^2\theta_r^*}{r^2\sin^2\theta_r^*}\nOmega^*\times\nr,
\end{equation}
corresponding to a rotational movement around the axis $\nOmega^*$,
where $\theta_r^*$ is the angle between $\nr$ and $\nOmega^*$.  The
angular dependence is determined by the geometry of the $d_{3/2}$
wave.  Going to momentum space, a dual expression can be obtained for
the expected value of the proton position for a given momentum
\begin{equation}\label{position}
\nr(\np) = -\frac{1+\sin^2\theta_p^*}{r^2\sin^2\theta_p^*}\nOmega^*\times\np,
\end{equation}
where now $\theta_p$ is the angle between $\nOmega^*$ and $\np$.  This
equation means that, given the orbit polarization direction
$\nOmega^*$ and the value of missing momentum $\np$ (determined by the
kinematics) one can compute the expected position of the
proton inside the orbit, as displayed in the examples of Fig.~4.
After the interaction with the virtual photon occurs, the proton exits the
nucleus with momentum $\np'=\np+\nq$. Since we know the initial
position of the proton inside the nucleus, we can determine the amount
of nuclear matter that the proton has to traverse before exiting,
and this in turn determines the total absorption due to the 
imaginary part of the
optical potential, {\it i.e.}, the reduction with respect to the PWIA cross
section. This effect was analyzed in depth in \cite{Ama99}.

In the case of nucleon recoil polarization of interest here, a second
ingredient is the expectation value of the proton spin for momentum $\np$
within the orbit, which is related to the spin density in momentum
space \cite{Ama02}. The corresponding spin-direction field for the
$d_{3/2}$ shell was computed in \cite{Ama02} and is given by
\begin{equation} \label{spin}
\ns(\np)=2\frac{\nOmega^*\cdot\np}{p^2}\np-\nOmega^*.
\end{equation}
In particular, if $\np$ is perpendicular to $\nOmega^*$, as in Fig.~4, then
$\ns=-\nOmega^*$, corresponding to spin opposite to the polarization
direction, as expected for a $d_{3/2}$ wave.  The above equation can
be inverted to obtain the orbit polarization as a function of the
proton spin:
\begin{equation}
\nOmega^*= 2\frac{\ns\cdot\np}{p^2}\np-\ns.
\end{equation}
Since the (detected) proton spin direction $\vec{s}$ is known for all of the
panels of Figs.~2 and 3, we can readily determine the corresponding orbit
polarization direction $\nOmega^*$ using Eq.~(\ref{spin}), and therefore, from
Eq.~(\ref{position}), also the proton position for each value of the missing
momentum.  This was done for the case of polarized $^{39}$K in
\cite{Ama99,Ama02}, where, from the nuclear transparencies computed as the
quotient between PWIA and DWIA cross sections, it was possible to extract a
theoretical value for the proton mean free path in our model.

As an example, in the simplest case of normal polarization we can have the two
situations depicted in Fig.~4. Both correspond to in-plane ($\phi=0$)
knockout of a proton for quasielastic kinematics; that is, $p\simeq q$ and
$\np$ is almost perpendicular to $\nq$ (quasi-perpendicular kinematics). In
both cases the proton is ejected from the $d_{3/2}$ shell of $^{40}$Ca, and
the corresponding orbit is represented by the shaded region as a planar
projection.  The only difference between the two cases is the direction of the
initial spin.  In the first case (top) the proton spin points upwards, with
$\theta_s=90^{\rm o}$ and $\Delta\phi=-\phi_s=-90^{\rm o}$, corresponding to
the panel in the second row and third column of Fig.~3.  Here the initial proton
is located in the half-plane with $z$ negative, because it is following the
orbit labeled ``orbit 1'' in Fig.~4 in clockwise sense.  Since it is exiting
the nucleus with momentum $\np'$, it has to cross the entire nucleus, which
implies large absorption due to FSI along its path. This is why in Fig.~3 the
FSI due to absorption is large (dotted lines).  In contrast, for the opposite
spin direction (bottom), $\Delta\phi=90^{\rm o}$, corresponding to the panel
in the third row and third column in Fig.~2, the proton is following 
``orbit 2'' in Fig.~4, in a counter-clockwise sense.  Therefore the initial
proton position is now in the positive $z$ region, while its final momentum is
the same, $\np'$, as before. Hence it has to cross only the nuclear surface
along its path, implying small absorption due to FSI, in accord with the
results of Fig.~2.

\begin{figure}[t]
\begin{center}
\leavevmode
\def\epsfsize#1#2{0.9#1}
\epsfbox{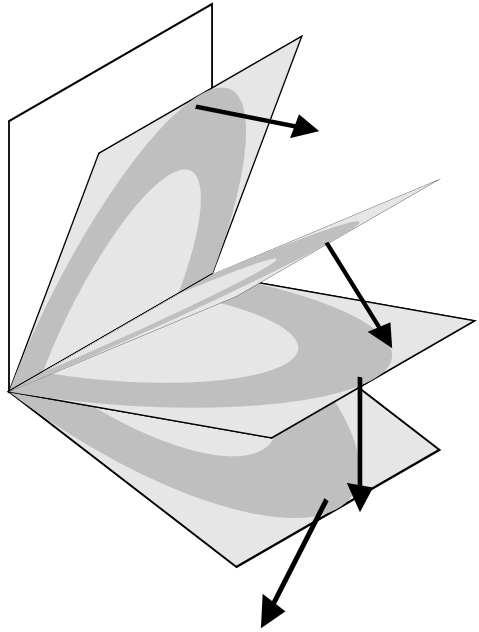}
\caption{Example of different orbits for which the initial proton has 
non-zero spin projection in the normal direction.}
\end{center}
\end{figure}

Thus the semi-classical picture that was able to explain FSI effects due to
absorption in $(e,e'p)$ reactions from polarized nuclei (at least in proton
knockout from the valence shell and for low missing momentum) is also
consistent with the case considered here of recoil polarization.  
This demonstration
is the main goal of this work and clearly provides a connection between the two
kinds of reactions, implying that both are consistent from this point of view.
We have shown that the mechanism that connects FSI with target polarization
in one reaction is the same one that induces recoil polarization in the other.
However there are also quantitative differences between the FSI effects in
the two processes. Namely, for the polarizations for which FSI is large (Fig.~3)
the effect is not as pronounced as in the case of a polarized nucleus (Fig.~1 of
\cite{Ama99}).  In other words the maximum reduction due to FSI is larger for the polarized target case than for the recoil polarization case.  The opposite happens for the cases where the FSI is small (Fig.~2), {\it i.e.}, the 
minimum reduction due to FSI
is smaller for the polarized target case than for the recoil polarization case 
(compare with Fig.~2 of \cite{Ama99}).

The reason for these differences is of a basic quantum nature related to 
the different kinds of measurements involved in the two reactions, since
in one case the polarization refers to the initial state, while in the other
it is the final state that is polarized.  Indeed, in the case of polarized
target the system is initially prepared in a state polarized in a given
direction (we assume 100\% polarization here); hence the orbit is uniquely
determined, as in the semi-classical picture in Fig.~4 for instance.  In the
case of recoil polarization, one measures the polarization of the proton in a
given direction $\vec{s}$ as the projection over the state $|\np
\vec{s}\rangle$; however, there are many spin directions that have non-zero
projections in the direction $\vec{s}$.  In other words, the final state is
not fully polarized in one direction before the measurement. This is, of course,
related to a fundamental postulate of quantum theory; only {\em after} the
measurement can one assure that the proton is polarized. Therefore within the
semi-classical picture, there are many possible initial orbits for the initial
proton with spin direction having a non-zero projection in the direction of
measurement, $\vec{s}$. In the example of Fig.~4, rotations of the initial
orbit, as shown schematically in Fig.~5, give a net spin projection in the
normal direction. Hence all of the orbits of Fig.~5 contribute to the final cross
section, with a weighting proportional to the cosine of the angle. Since these
orbits imply different initial positions for the proton, and therefore
different magnitudes for the FSI effects, the final cross section that 
results is an average
between cases where FSI is large and other where it is not so large.
Accordingly, the magnitude of FSI in the results of Figs.~2 and 3 can be
understood as average values coming from the different polarizations in the
final state, or as average values of the FSI effects found in the polarized
target case of \cite{Ama99}.

We should add that this implies a fundamental difference between polarized
target and recoil polarization reactions. In the first case the net
polarization could reach 100\%, at least theoretically.  In the second
case  is not possible to have 100\% polarization in the final state before
the measurement. In this sense, the physical information that can be obtained
in reactions with polarized targets will be always richer than that obtained
with the recoil polarization reactions that unavoidably contain polarization averages.
 
\begin{figure}[tb]
\begin{center}

\leavevmode
\def\epsfsize#1#2{0.8#1}
\epsfbox{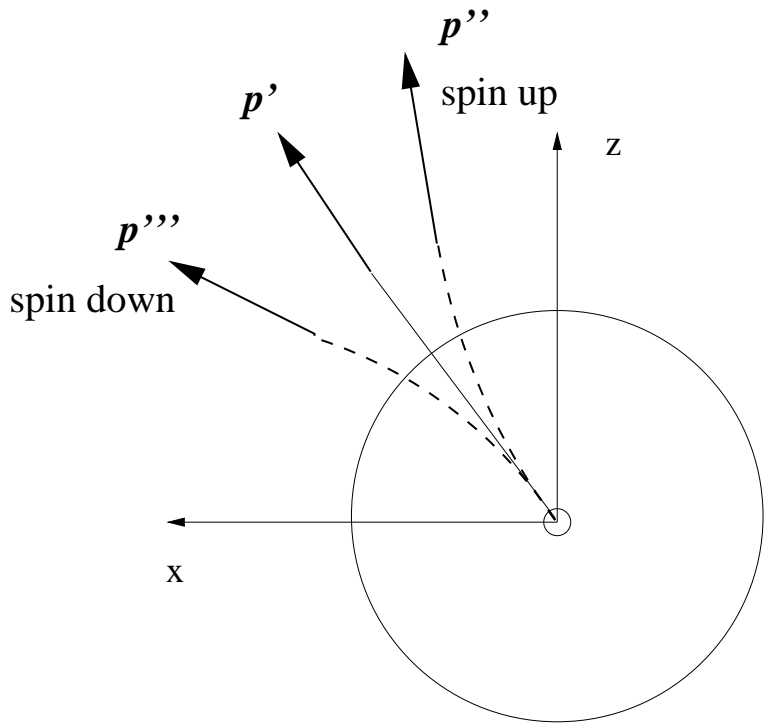}
\caption{Deviations of ejected proton trajectories due to the real part of the
  spin-orbit potential for spin-up and spin-down polarizations (see Fig.~4 for
  kinematics).}
\end{center}
\end{figure}

\begin{figure}[tb]
\begin{center}
\leavevmode
\def\epsfsize#1#2{0.9#1}
\epsfbox[150 580 450 780]{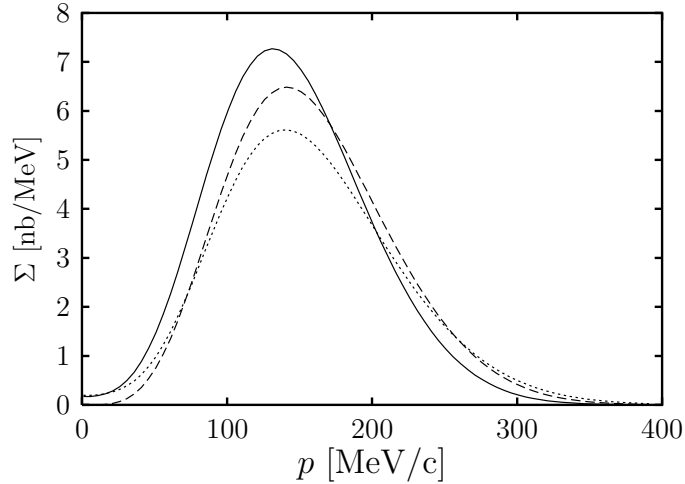}
\caption{Cross section for proton knockout from the $d_{3/2}$ shell
of $^{40}$Ca with electron scattering angle $\theta_e=30^{\rm o}$ and
$\phi=0$.  The DWIA calculations include only the real part of the
spin-orbit in the optical potential for spin-up (normal direction ${\vec n}$,
solid lines) and spin-down ($-{\vec n}$ direction, dotted lines) polarized
protons. The PWIA is shown with dashed lines. The spin-up
distribution is shifted to the left of the spin-down one.}
\end{center}
\end{figure}

Up to now we have centered our attention on the imaginary part of the
optical potential which gives the major effect for the cross section
results of Figs.~2 and 3. The remaining effect is mainly produced by
the real part of the spin-orbit potential.  In \cite{Woo98} a
semi-classical argument was used to justify the sign of the induced
polarization measured in $^{12}$C. Following \cite{Woo98} the  spin-orbit 
force can be written
\begin{equation} \label{force}
\nF = -\nabla\left[V_{ls}(r)\nsigma\cdot\nL\right]
= -\hat{\nr}\frac{d V_{ls}}{dr}  \nsigma\cdot\nL
+ V_{ls}\nsigma\times\np'.
\end{equation}
Let us consider the case of $\phi=0$ and normal polarization ${\vec s} = \pm
\vec{n}$.  Assuming $V_{ls}<0$ and constant, the effect of the second
term in Eq.~(\ref{force}) is to rotate the trajectory, with a force $\nF
\propto \np'\times\vec{s}$. This is shown schematically in Fig.~6,
where a nucleon is emitted with final momentum $\np'$, and spin
polarization pointing in the direction normal to the plane of the
figure. The final momentum changes due to the spin-orbit force, in
such a way that nucleons with spin up (pointing out of the plane) are
shifted to its right, {\it i.e.}, to smaller angles from $\nq$, while
nucleons with spin down (pointing into the plane) are shifted to its
left, {\it i.e.}, to larger angles.  In other words, the missing momentum of
nucleons with spin up decreases, while that of nucleons with spin down
increases, and hence one should expect a shift of the missing momentum
distribution of nucleons with spin up to the left of that for nucleons
with spin down.  

Such a shift can be verified from our DWIA results in Fig.~7, where we
show the total cross section for protons polarized in the $\pm
\vec{n}$ directions for the same kinematics as Figs.~2 and 3, but
including only the real part of the spin-orbit in the optical
potential.  The solid lines correspond to normal up polarization, and
the dotted lines to the opposite polarization. In the figure we can
see the shift between the two distributions as suggested by the
semi-classical argument.  However, it is also clear that the difference
between the two polarizations is not only a shift, but also that the maximum
of the cross section for $\vec{n}$ polarization is bigger than for
$-\vec{n}$ polarization. This should be related to the first term
neglected in Eq.~(\ref{force}) that goes to increase or decrease the
final momentum depending on the polarization and on the gradient of
the spin-orbit potential. The net effect is not easy to depict in
geometrical terms.  In the figure we also show for comparison the PWIA
cross section (dashed lines), which does not depend on polarization.
Moreover, upon comparing with these PWIA results, we see that only the
spin-up distribution is clearly shifted (to the left), while the position of
the spin-down distribution is basically the same as the PWIA one.

\begin{figure}[tp]
\begin{center}
\leavevmode
\def\epsfsize#1#2{0.9#1}
\epsfbox[100 400 500 810]{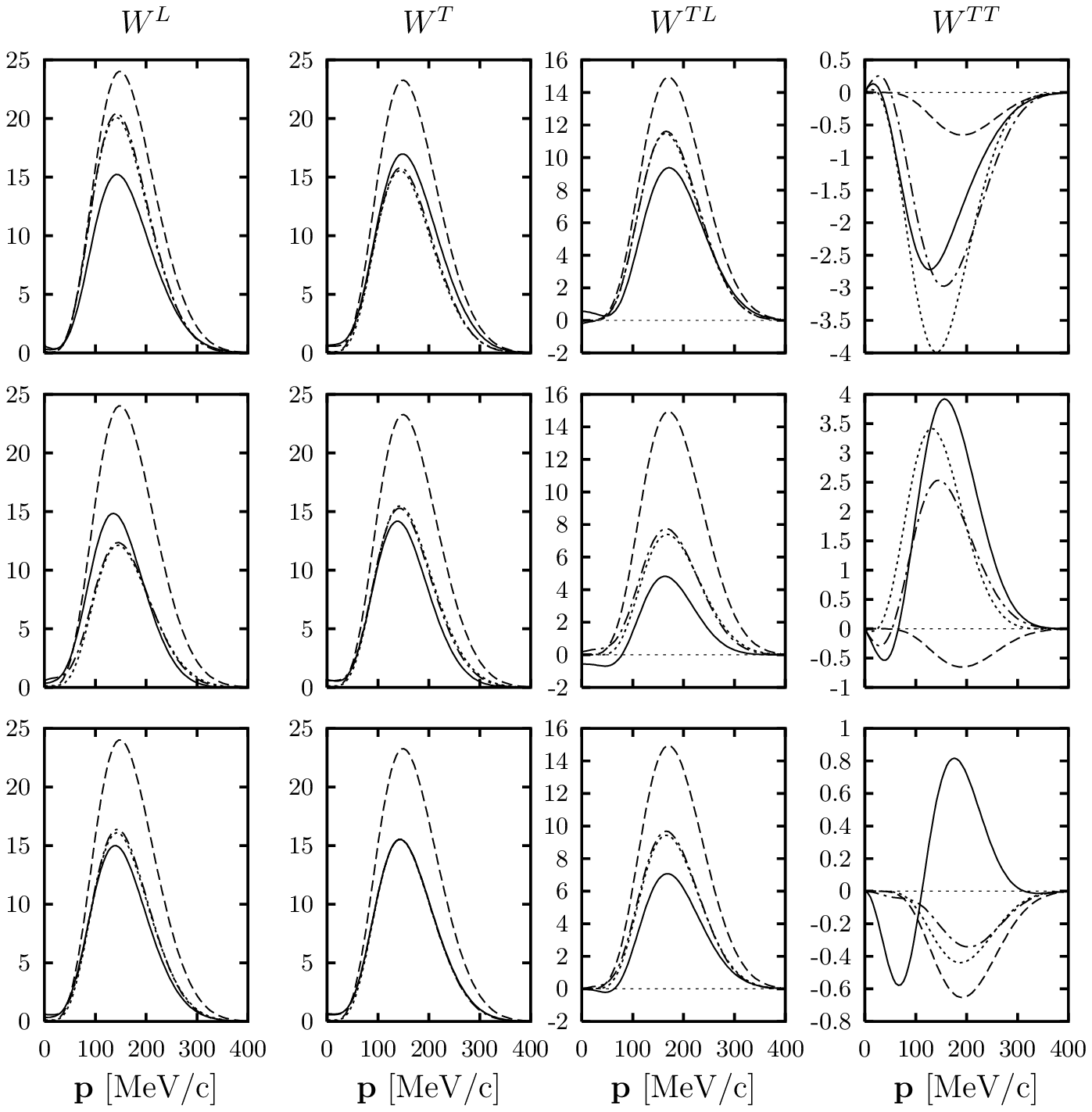}
\caption{Separate response functions in units of fm$^ 3$ for proton
knockout from the $d_{3/2}$ shell of $^{40}$Ca. The kinematics are the
same as in Figs.~2 and 3. We show results corresponding to two
different polarization directions for the recoil proton. Upper panels:
proton polarized in the $-\vec{n}$ direction ($\theta_s=90^{\rm o}$,
$\Delta\phi=90^{\rm o}$).  Middle panels: polarized in the $\vec{n}$
direction ($\theta_s=90^{\rm o}$, $\Delta\phi=-90^{\rm o}$).
Lower panels: polarized in the $z$-direction, $\theta_s=0^{\rm o}$.  }
\end{center}
\end{figure}

\begin{figure}[tb]
\begin{center}
\leavevmode
\def\epsfsize#1#2{0.9#1}
\epsfbox[100 650 500 810]{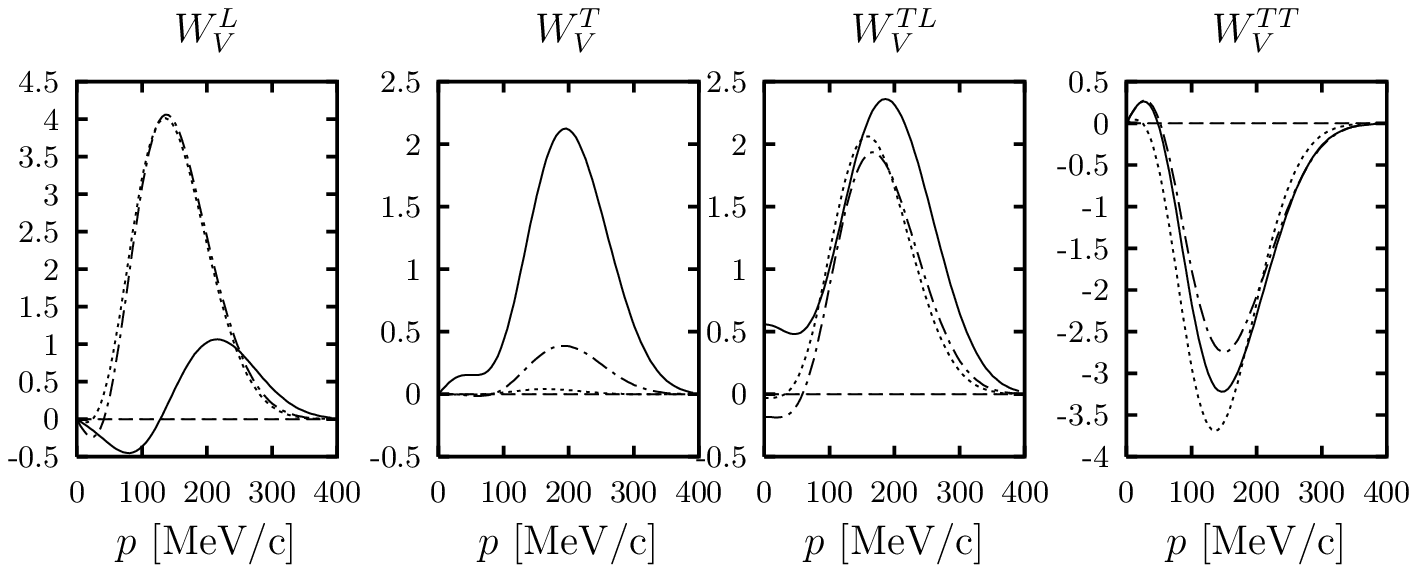}
\caption{Vector parts of the separate response functions 
of Fig.~8 for $-{\vec n}$ polarization}
\end{center}
\end{figure}

\subsection{Other polarization observables}

To complete our discussion, in Figs.~8 and 9 we present results for
observables other than the cross section. In Fig.~8 the separate response
functions contributing to the cross section of Figs.~2 and 3 are shown for
three directions of proton polarization: the $-\vec{n}$ 
polarization (upper panels),
corresponding to the panel $\theta_s=90^{\rm o}$ and $\Delta\phi=90^{\rm o}$
of Fig.~2; $\vec{n}$ polarization (middle panels) corresponding to the panel
$\theta_s=90^{\rm o}$ and $\Delta\phi=270^{\rm o}$ of Fig.~3; and finally
polarization in the $z$ direction (lower panels), corresponding to the case
$\theta_s=0$ of Fig.~2.  Since the azimuthal angle is $\phi=0$, the responses
$\widetilde{W}^{TL}$ and $\widetilde{W}^{TT}$ do not contribute to this cross
section (see Eqs.~(\ref{RL}--\ref{RTT})). For the same reason, the responses
for $z$-polarization and $\phi=0$ do not have contributions from the
spin-vector part (see Eq.~(\ref{WK})) and are identical to the unpolarized
responses except for a factor $1/2$.  Figure~8 is complemented with Fig.~9,
where for $-\vec{n}$ polarization we show the vector part $W^K_V$ of 
the response
functions of Fig.~8 (see Eq.~(\ref{WK})). This contribution 
is precisely the difference between the upper and lower panels of Fig.~8,
and also the difference between the lower and middle panels. 
These responses could in principle be extracted from experiment by performing
a super-Rosenbluth plot for different kinematics.

The results of Fig.~8 illustrate the varied effects that the  FSI have on the
different responses when viewed as functions of the polarization.  First we note
that differences already exist between the L and T responses.  For the L
response the effect of the central imaginary part of the optical potential,
$W_C$, is similar to the one found for the cross section: it is a small
reduction for $-\vec{n}$ polarization and large for 
$\vec{n}$ polarization. However, in the
case of the T response, this effect is basically the same for all of the
polarizations.  The reason can be deduced by examining Fig.~9, where
 $W^T_V$ is almost
zero if only $W_C$ is included, while $W^L_V$ is large in that case.  This
different behavior is related to the fact that the vector responses are zero
in the absence of FSI, and to the different spin dependence of the L and T
components of the electromagnetic current operator.  The effect of $W_C$ on
the TL response, $W^{TL}$, has a similar behavior to that on the L response, being a small reduction for $-\vec{n}$ polarization, and large for 
$\vec{n}$ polarization 
(Fig.~8).  This is because the vector response $W^{TL}_V$ is approximately
proportional to $W^L_V$, with a factor 1/2, when only the $W_C$ potential is
used in the calculation (Fig.~9).  The case of the TT response is different from
the others because in PWIA this response is very small compared with them (in
fact, it is exactly zero to first order in an expansion in powers of $p/m_N$).
Moreover, the unpolarized TT response (bottom panel in Fig.~8) is still
small when FSI are included. However, the effect of FSI is amplified in the
polarized case when we approach the $\pm \vec{n}$ polarization.  
The size of this
response changes dramatically (almost an order of magnitude) when the optical
potential is included, and this effect is dominated by the imaginary central
part, $W_C$. In fact, from Fig.~9, we see that the response $W^{TT}_V$ for
$-\vec{n}$ polarization is of the same magnitude as the other three vector
responses, or even larger when the full FSI is included.
Concerning the real part of the central optical potential, it has a very small
effect on these observables. 

\begin{figure}[tb]
\begin{center}
\leavevmode
\def\epsfsize#1#2{0.9#1}
\epsfbox[90 610 480 780]{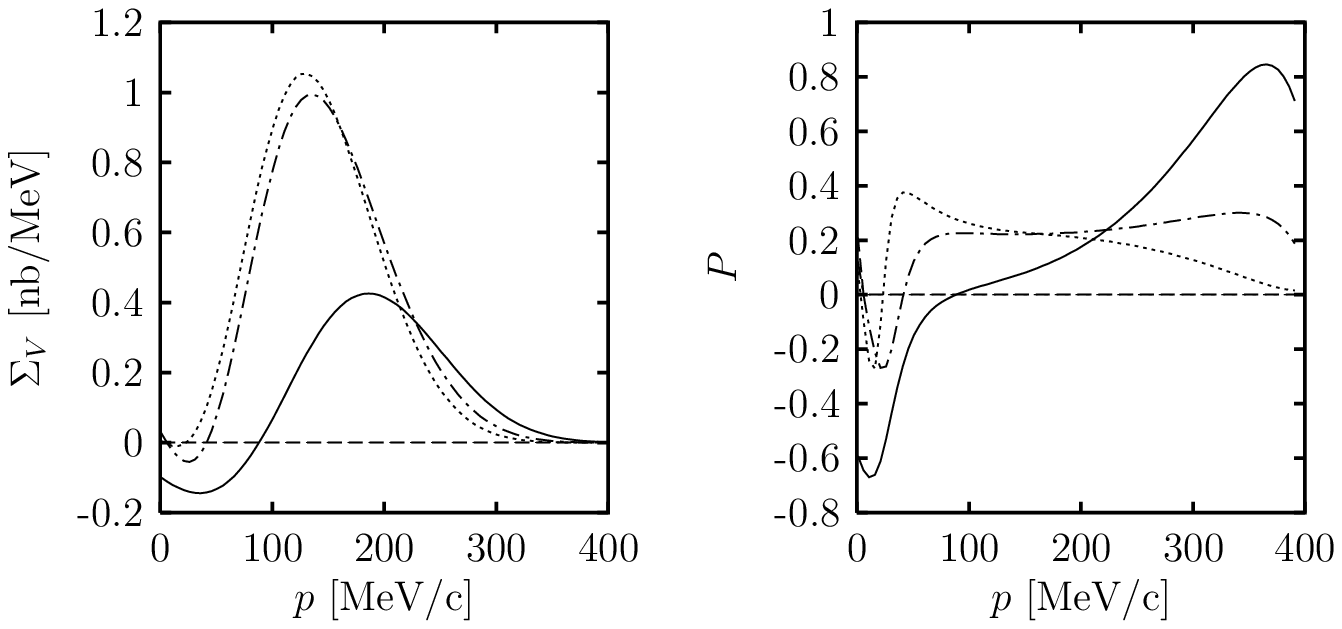}
\caption{
Vector part of the cross section and induced polarization 
in the $-{\vec n}$ direction for the same kinematics as in Fig.~2.}
\end{center}
\end{figure}

The spin-orbit potential has a different impact.  It produces a large reduction
of $W^L_V$ (Fig.~9) in such a way that this vector response is close to zero
near the peak of the momentum distribution, and hence the final polarized L
response changes very slightly with the polarization angles (Fig.~8).  The
spin-orbit potential also has a large effect on the $W^T_V$ response 
(Fig.~9), which is almost zero in absence of $V_{ls}$. 
On the other hand, the contribution of $V_{ls}$
to the $W^{TL}$ response is already large 
in the unpolarized case; however, this potential 
provides a small contribution to $W^{TL}_V$.  The net effect of FSI on the
$T$ and $TL$ responses is to produce a reduction which is smaller 
for $-\vec{n}$ than for $\vec{n}$
polarization.  Finally, a very large effect of FSI is seen in the TT
response, which even changes sign in going from 
$-\vec{n}$ to $\vec{n}$ polarization. The
spin-orbit potential also produces a small effect on the $W^{TT}_V$
response.  All of the various effects seen in Figs.~8 and 9 coming from FSI
interplay to produce the results of Figs.~2 and 3.

To finish, in Fig.~10 we show for the same kinematics other
observables of interest that can be extracted from the results of
Figs.~2 and 3.  The left-hand panel contains the vector part of the
cross section, $\Sigma_V$, for $-\vec{n}$ polarization. This can be
obtained as the difference between the corresponding cross section for
$-\vec{n}$ polarization and the upper panel in Fig.~2.  In the
right-hand panel we show the induced polarization also in the
$-\vec{n}$ direction.  In this figure we can see again the
contribution of the central parts of the potential and the full DWIA
calculation.  The spin-orbit potential reduces both $\Sigma_V$ and $P$
for low and intermediate missing momentum and increases the induced
polarization for high $p$.  The large values of the induced
polarization for high missing momentum ($P\sim 0.8$ for $p>300$ MeV/c)
has also been predicted in previous work \cite{Woo98,Udi00,Kaz04}, but
in this region other reaction mechanisms are also important, such as
MEC or dynamical relativistic effects.


\section{Conclusions}


In this work we have performed a thorough study of polarization
observables in exclusive $(e,e'\vec{p})$ reactions leading to discrete
residual nucleus states.  We have used a semi-relativistic DWIA model
including relativistic corrections to the one-body current as well as
relativistic kinematics. Our model was succesfully confronted with the
available experimental data for polarization observables in
\cite{Kaz03,Kaz04}.  In the present work we have applied 
our approach to proton
knockout from the outer shell of $^{40}$Ca for quasielastic
kinematics and in-plane emission.  We have focused on the
dependence of the final-state interaction on the polarization
angles. In this way we have been able to identify the polarization
directions for which the FSI effects are large or small. This dependence has
been compared with the findings of \cite{Ama99} for a polarized
$^{39}$K nucleus, where the effects arising from the central imaginary part
of the optical potential were explained using a semi-classical model
based on the location of the proton in the orbit.  

In this paper we
have shown that the same semi-classical model is also able to explain
the results of recoil polarization, using the relationship between
polarization of the orbit and expectation value of the spin.  However, we
have also found quantitative differences between FSI for the two
reactions.  For recoil polarization, the FSI effects 
present a softer dependence on the polarization angles. 
An explanation for this fact has been proposed
by calling on the measurement postulate of quantum mechanics: in the
case of recoil polarization the final proton is not really polarized
before the measurement.  Hence the initial orbit of the proton cannot
be completely fixed, since there are other possible orientations
giving a projection of the proton spin on the measurement direction,
and all of these orbits contribute in a different way to the FSI.
Thus the semi-classical orbit only determines the 
 most probable orientation of the proton.

We have ended this paper by analyzing the separate response functions and
other polarization observables for selected polarization angles,
emphasizing the different sensitivities to the various contributions of 
the FSI. The case of the spin-vector part of the response functions
for normal polarization is especially instructive, since these
observables exhibit the strongest influences from the interaction, and
hence their determination would place strong constraints on any
theoretical calculation.

Differences between polarized target and recoil polarization
arise irrespective of the underlying dynamics used. This is
reminiscent of what is seen for the various polarization observables that contribute
to the two types of reactions. In this sense, these two reactions are
complementary to one another, and both are  needed if one wants to
extract all of the information available in the general case
of exclusive reactions.

The unified picture that emerges from the present comparison between
polarized target and recoil polarization is a good starting point to
understand in an intuitive way how the underlying nuclear dynamics
interact with the polarization degrees of freedom and affect the cross
section for different polarizations. Thus, even if the complexity of
full sets of polarization observables has not stimulated more
experimental and theoretical research, we hope that the comprehensive
view that we are reaching about polarization observables in these
reactions encourage further studies in this field.

\section*{Acknowledgments}
This work was partially supported by funds provided by DGI (Spain) and
FEDER funds, under Contracts Nos BFM2002-03218, BFM2002-03315 
and FPA2002-04181-C04-04 and
by the Junta de Andaluc\'{\i}a. It was also supported in part (TWD) by
the U.S. Department of Energy under cooperative agreement No.
DE-FC02-94ER40818.



\begin{thebibliography}{exp92}


\bibitem{Woo98}  R.J. Woo {\em et al.},
                 Phys. Rev. Lett. {\bf 80} (1998) 456.

\bibitem{Mal00}  S. Malov {\em et al.},
                 Phys. Rev. C {\bf 62} (2000) 057302.

\bibitem{Kel96}  J. J. Kelly, 
                 Adv. Nucl. Phys. {\bf 23} (1996) 75.

\bibitem{Bof96}  S. Boffi, C. Giusti, F. D. Pacati, and M. Radici,
                 {\em Electromagnetic response of atomic nuclei},
                 Oxford University Press (1996).

\bibitem{Ito97}  H. Ito, S.E. Koonin, and R. Seki, 
                 Phys. Rev. C {\bf 56} (1997) 3231.

\bibitem{Ryc99}  J. Ryckebusch, D. Debruyne, W. Van Nespen, and S. Janssen,
                 Phys. Rev. C {\bf 60} (1999) 034604.

\bibitem{Kel99}  J.J. Kelly, 
                 Phys. Rev. C {\bf 59} (1999) 3256.

\bibitem{Kel99b} J.J. Kelly, 
                 Phys. Rev. C {\bf 60} (1999) 044609.

\bibitem{Joh99}  J.I. Johanson and H.S. Sherif,
                 Phys. Rev. C {\bf 59} (1999) 3481.

\bibitem{Udi00}  J.M. Ud\'{\i}as and J.R. Vignote, 
                 Phys. Rev. C {\bf 62} (2000) 034302.

\bibitem{Kaz03}  F. Kazemi Tabatabaei, J.E. Amaro, and J.A. Caballero,
                 Phys. Rev. C {\bf 68} (2003) 034611.

\bibitem{Kaz04}  F. Kazemi Tabatabaei, J.E. Amaro, and J.A. Caballero,
                 Phys. Rev. C {\bf 69} (2004) 064607.

\bibitem{Mar02a} M.C. Mart\'{\i}nez, J.A. Caballero, and T.W. Donnelly,
                 Nucl. Phys. A {\bf 707} (2002) 83.

\bibitem{Mar02b} M.C. Mart\'{\i}nez, J.A. Caballero, and T.W. Donnelly,
                 Nucl. Phys. A {\bf 707} (2002) 121.

\bibitem{Mar03} M.C. Mart\'{\i}nez, J.R. Vignote, J.A. Caballero, T.W. Donnelly,
                E. Moya de Guerra, and J.M. Ud\'{\i}as,
                Phys. Rev. C {\bf 69} (2004) 034604.

\bibitem{Vig03} J.R. Vignote, M.C. Mart\'{\i}nez, J.A. Caballero, 
                E. Moya de Guerra, and J.M. Ud\'{\i}as,
                Phys. Rev. C {\bf 70} (2004) 044608.

\bibitem{Die01} S. Dieterich {\em et al.}, 
                Phys. Lett. B {\bf 500} (2001) 47.

\bibitem{Str03} S. Strauch {\em et al.,} 
                Phys. Rev. Lett. {\bf 91} (2003) 052301.

\bibitem{Fru84}  S. Frullani and J. Mougey, 
                 Adv. Nucl. Phys. {\bf 14} (1984) 1.

\bibitem{Ras89}  A.S. Raskin and T.W. Donnelly,
                 Ann. Phys. (N.Y.) {\bf 191} (1989) 78.

\bibitem{Bof93}  S. Boffi, C. Giusti, and F.D. Pacati, 
                 Phys. Rep. {\bf 226} (1993) 1.

\bibitem{New53} H.C. News,
                Proc. Phys. Soc. London Sect. A {\bf 66} (1953) 477.

\bibitem{New58} H.C. News and M.Y. Refai,
                Proc. Phys. Soc. London Sect. A {\bf 71} (1958) 627.

\bibitem{Jac76} G. Jacob, Th. A.J. Maris, C. Schneider, and
                M.R. Teodoro,
                Nucl. Phys. A {\bf 257} (1976) 517.

\bibitem{Ama99} J.E. Amaro and T.W. Donnelly,
                Nucl. Phys. A {\bf 646} (1999) 187.

\bibitem{Ama02} J.E. Amaro and T.W. Donnelly,
                Nucl. Phys. A {\bf 703} (2002) 541.

\bibitem{Cab04} J.E. Amaro, M.B. Barbaro, and J.A. Caballero, 
                nucl-th/0411043.

\bibitem{Are04} H. Arenh\"ovel, W. Leidemann, and E.L. Tomusiak,
                nucl-th/0407053.

\bibitem{Are02} H. Arenh\"ovel, W. Leidemann, and E.L. Tomusiak,
                    Eur. Phys. Jou. A (2002) 491.

\bibitem{Are00} H. Arenh\"ovel, W. Leidemann, and E.L. Tomusiak,
                  Few Body Syst. {\bf 28} (2000) 147.

\bibitem{Are98} H. Arenh\"ovel, W. Leidemann, and E.L. Tomusiak,
                 Nucl.Phys. A {\bf 641} (1998) 517.

\bibitem{Are95} H. Arenh\"ovel, W. Leidemann, and E.L. Tomusiak,
                Phys. Rev. C {\bf 52} (1995) 1232.

\bibitem{Are92} H. Arenh\"ovel, W. Leidemann, and E.L. Tomusiak,
               Phys. Rec C {\bf 46} (1992) 455.

\bibitem{Dmi89} V. Dmitrasinovic and F. Gross,
                Phys. Rev. C {\bf 40} (1989) 2479.

\bibitem{Ama03b} J.E. Amaro, M.B. Barbaro, J.A. Caballero, and
                 F. Kazemi Tabatabaei,
                 Phys. Rev. C {\bf 68} (2003) 014604.

\bibitem{Ama98b} J.E. Amaro and T.W. Donnelly,
                 Ann. Phys. (N.Y.) {\bf 263} (1998) 56.

\bibitem{Pic87}  A. Picklesimer and J.W. Van Orden,
                 Phys. Rev. C {\bf 35} (1987) 266.

\bibitem{Pic89}  A. Picklesimer and J.W. Van Orden,
                 Phys. Rev. C {\bf 40} (1989) 290.

\bibitem{Maz02}  M. Mazziotta, J.E. Amaro, and F. Arias de Saavedra,
                 Phys. Rev. C {\bf 65} (2002) 034602.

\bibitem{Sch82}  P. Schwandt {\em et. al.,}
                 Phys. Rev. C {\bf 26} (1982) 55.

\bibitem {Udi99} J. M. Ud\'{\i}as, J. A. Caballero, E. Moya de Guerra, 
                 J. E. Amaro, and T. W. Donnelly, 
                 Phys. Rev. Lett. {\bf 83} (1999) 5451. 

\bibitem{Ama96a} J.E. Amaro, J.A. Caballero, T.W. Donnelly, 
                 E. Moya de Guerra, A.M. Lallena, and J.M. Ud\'{\i}as, 
                 Nucl. Phys. A {\bf 602} (1996) 263.

\bibitem{Ama98a} J.E. Amaro, M.B. Barbaro, J.A. Caballero, 
                 T.W. Donnelly, and A. Molinari,
                 Nucl. Phys. A {\bf 643} (1998) 349.

\bibitem{Ama02c} J.E. Amaro, M.B. Barbaro, J.A. Caballero, 
                 T.W. Donnelly, and A. Molinari,
                 Phys. Rep. {\bf 368} (2002) 317.

\bibitem{Ama03} J.E. Amaro, M.B. Barbaro, J.A. Caballero, 
                 T.W. Donnelly, and A. Molinari,
                 Nucl. Phys. A {\bf 723} (2003) 181.


\end{thebibliography}
\end{document}